\documentclass[12pt]{article}
\usepackage{amsmath}
\usepackage{graphicx}
\usepackage{enumerate}
\usepackage{natbib}
\usepackage{titlesec}
\titlelabel{\thetitle.\quad}
 
\usepackage{tikz}
\usetikzlibrary{matrix}
\usepackage{makecell}
\usepackage{caption}
\usepackage{float}
\usepackage{tablefootnote}

\newcommand{\blind}{0}

\date{}
\begin{document}

\def\spacingset#1{\renewcommand{\baselinestretch}%
{#1}\small\normalsize} \spacingset{1}


\if0\blind
{
  \title{\bf Weighted Approach for Estimating Effects in Principal Strata with Missing Data for a Categorical Post-Baseline Variable in Randomized Controlled Trials}
  \author{Shengchun Kong\\
    \it{Genentech, South San Francisco, CA 94080}\\
	\and\\
    Dominik Heinzmann \\
	\it{F. Hoffmann-La Roche Ltd., 4070 Basel, Switzerland}\\
	\and\\
    Sabine Lauer\\
	\it{Dr. Lauer Research, 63263 Neu-Isenburg, Germany}\\
	\and\\
    Tian Lu\\
	\it{Stanford University, Palo Alto, CA 94305}}
  \maketitle
} \fi

\if1\blind
{
  \bigskip
  \bigskip
  \bigskip
  \begin{center}
    {\LARGE\bf Weighted Approach for Principal Stratum Incorporating Missing Data for Categorical Post-Baseline Variables in Randomized Controlled Trials}
\end{center}
  \medskip
} \fi

\bigskip
\begin{abstract}
This research was motivated by studying anti-drug antibody (ADA) formation and its potential impact on long-term benefit of a biologic treatment in a randomized controlled trial, in which ADA status was not only unobserved in the control arm but also in a subset of patients from the experimental treatment arm. Recent literature considers the principal stratum estimand strategy to estimate treatment effect in groups of patients defined by an intercurrent status, i.e. in groups defined by a post-randomization variable only observed in one arm and potentially associated with the outcome. However, status information might be missing even for a non-negligible number of patients in the experimental arm. For this setting, a novel weighted principal stratum approach is presented: Data from patients with missing intercurrent event status were re-weighted based on baseline covariates and additional longitudinal information. A theoretical justification of the proposed approach is provided for different types of outcomes, and assumptions allowing for causal conclusions on treatment effect are specified and investigated. Simulations demonstrated that the proposed method yielded valid inference and was robust against certain violations of assumptions. The method was shown to perform well in a clinical study with ADA status as an intercurrent event.
\end{abstract}

\noindent%
\vfill

\newpage
\spacingset{1.5} 
\section{INTRODUCTION}
\label{s:intro}

Many clinical variables can change in a patient after initiation of a therapeutic treatment and can help identify patients who are more likely to respond favorably to the treatment in terms of long-term outcomes such as overall survival. In a randomized controlled trial (RCT), often such post-randomization variables are only observed in the experimental treatment arm. Examples are adherence to or discontinuation from the experimental treatment, and biomarkers such as treatment-specific adverse reactions, pharmacokinetic measures, or anti-drug antibodies (ADAs) for biologic therapies. In those settings, if the post-randomization variable $A$ can reliably be organized into categories $A=1,\cdots,J$, two important scientific questions of interest are (i) is the treatment benefit clinically meaningful for all categories of $A$ and (ii) is the treatment effect similar between certain or all categories of $A$. 

In our motivating example (see Section \ref{sec_ada}), we observed that Atezolizumab treated patients with positive ADA-status had worse outcome for overall survival compared to patients with negative ADA-status. However, ADA-positive patients also showed differences in several prognostic baseline variables compared to ADA-negative patients. Consequently, the observed difference in outcome must not necessarily translate into a diminished treatment effect in ADA-positive patients compared to ADA-negative patients because the adequate (unobserved) subgroups of patients in the control arm (patients who would/would not develop ADAs if they were treated with Atezolizumab) should also differ in prognosis and thus outcome. A naive comparison with the full control arm for both ADA subgroups simply reflects the differences in outcome between the ADA subgroups in the experimental arms and leads to biased treatment effect estimates. It does not allow to causally assess the impact of ADA-status on treatment effect.

The determination of treatment effect in groups of patients defined through a post randomization categorical variable in the experimental treatment group is not straight forward \citep{Bornkamp:2019}. If the categorical post-randomization variable $A$ is observed after treatment initiation and either precludes observation or affects interpretation of the clinical endpoint of interest, it can be considered an intercurrent event in the language of the International Council for Harmonization (ICH) E9 addendum \citep{ICHE9:2019}. 

The principal stratum estimand strategy, originally introduced by \cite{Frangakis:2002}, can be considered to address the above scientific questions of interest. A comprehensive review of possible assumptions for principal stratum estimands is provided in \citet{Ding:2017}. Recent applications of the principal stratum strategy for a post-baseline binary biomarker and for treatment compliers are provided in \citet{Bornkamp:2019} and in \citet{Larsen:2019}, respectively - each with a different set of assumptions to make the target estimand identifiable. \citet{Bornkamp:2019} considered a similar approach as \citet{Stuart:2015}, with one of the core assumptions being the conditional independence of outcome in the control arm and the post-randomization stratum variable given the set of covariates. In contrast, \citet{Larsen:2019} required that the outcome in the control arm be conditionally independent of the set of covariates given the post-randomization stratum variable. This assumption, however, implies that the underlying post-randomization stratum variable is the main driver of outcome in the control arm and, thus, differences in baseline covariates would have no influence on outcome within a given underlying stratum. The plausibility of such an assumption would likely decrease in the presence of well-known prognostic factors such as baseline disease staging, whose effect on the outcome needs to be fully explained by the post-randomization variable. 
This makes the assumption implausible for many applications, including the one considered in this paper. 

\citet{Larsen:2019} requires the post-baseline stratum to be fully observed in the experimental arm without missing data and \citet{Bornkamp:2019} only shortly refers to multiple imputation for missing biomarker status in an appendix without detailed discussions on the methodology and the assumptions needed. In many clinical trials including post-baseline measurements of a biomarkers, a non-negligible amount of missing data is present. Complete case analysis is not applicable because generally one would not observe those patients in the control arm who would have missing stratum information if they would have been treated with the experimental treatment (counterfactual missingness). Simple imputation to a fixed value for all patients with a missing stratum is also not optimal because it requires the assumptions that all patients with missing stratum information belong to the same stratum to yield valid causal inference. 

We consider the missing post-baseline principal stratum problem in this article and propose a novel weighted approach. Our approach degenerates to the Weight Placebo Patients (WPP) approach used in \citet{Bornkamp:2019} where  the post-baseline strata are observed for every subject. Our approach is illustrated via simulations as well as an application to ADAs of CIT in a pivotal RCT investigating the CIT atezolizumab in chemotherapy-naïve patients with metastatic non-squamous non-small cell lung cancer \citep{Socinski:2018}.


\section{METHODS} \label{smethod}

\subsection{Estimand of Interest} \label{ss:estimand}

We considered a general model framework to estimate the treatment effect in subgroups defined by post-treatment categorical variable in an RCT, in which the variable is observable in the experimental treatment arm, but not the other arm. These post-treatment variables may be longitudinally measured and are often internal time-varying variables that contain mortality information which could complicate the interpretation \citep{Kalbfleisch:2002}. Hence, survival models incorporating time-dependent external covariates are usually not appropriate here. A landmark approach, in which the subgroups are defined by the variable value at the landmark time point, is considered for this problem. Subjects with an event or censored before the landmark timepoint are excluded from the analyses. Considerations for the selection of a landmark time point are provided in the real-data example Appendix B and also discussed in \cite{Mattei:2020} and \cite{Bornkamp:2020}. 

The following notation is used: $R$ represents the treatment assignment of a particular individual, in which $R=1$ denotes the experimental treatment group and $R=0$ denotes the control group. Let $T=T^{(1)}R+T^{(0)}(1-R)$ be the observed outcome for a given individual, in which $T^{(1)}$  and $T^{(0)}$ are the potential outcomes if the individual is assigned to the experimental and control arms, respectively. Let $A^{(1)}$ represent the underlying post-randomization landmark subgroup stratum if the patient is receiving the experimental treatment. It allows up to $J$ levels. For individuals in the experimental arm, $A^{(1)}$ is generally observed. For individuals in the control arm, $A^{(1)}$ is unobservable by definition. However, one can assign a counterfactual status to those individuals assuming the stratum that they would be in if they had been treated with the experimental treatment. A simplified notation, $A$, is used in the following sections. 

Patients in the experimental treatment group may have a missing value for $A$ which could, at least partly, be related to outcome (e.g., biomarker assessment could not be conducted because blood could not be drawn from a patient due to their physical condition at the visits prior to the landmark time point). A simplified notation $M$, instead of counterfactual notation $M^{(1)}$, denotes an indicator variable for the missing value for $A$, with $M = 0$ and $1$ denoting a non-missing and missing value for $A$, respectively.  This notation will also be used for control arm patients to indicate the counterfactual “missing” category, i.e., those patients in the control group who would have a missing value for $A$ if they received the experimental treatment. Let $ {\cal E}_{Oa}$ and ${\cal E}_{Ma}$ be the groups of patients in the experimental arm with true (underlying) stratum $A=a$ for $M=0$ and $M=1,$ respectively. Analogously,  ${\cal C}_{Oa}$ and ${\cal C}_{Ma}$ denote groups of patients in the control arm with counterfactual stratum $A=a$ for $M=0$ and $M=1,$ respectively.  The setup is illustrated in Table \ref{tableA}. Finally, let $X$ denote the set of  covariates at baseline/landmark that influences $A$ and the clinical outcome and assume that $X$ is fully observed. 

\begin{table}
	\caption{Observed and counterfactual categories of individuals}\label{tableA}
	\begin{center}
		\begin{tabular}{ |c|c|c|c|c|c|c| } 		
			\hline
			& \multicolumn{3}{|c|}{$M=0$} & \multicolumn{3}{|c|}{$M=1$}\\
			\hline
			& $A=1$ & $\dots$ & $A=J$ & $A=1$ & $\dots$ & $A=J$\\
			\hline
			Experimental treatment & ${\cal E}_{O1}$ & $\dots$ &${\cal E}_{OJ}$ &${\cal E}_{M1}$&$\dots$&${\cal E}_{MJ}$\\
			\hline
			Control & ${\cal C}_{O1}$ &$\dots$ &${\cal C}_{OJ}$ &${\cal C}_{M1}$&$\dots$&${\cal C}_{MJ}$\\
			\hline
		\end{tabular}
	\end{center}
\end{table}

If the post-randomization variable $A$  has the potential to impact the interpretation of the clinical outcomes then it is an intercurrent event in the language of the ICH E9 addendum \citep{ICHE9:2019}. Hence, the principal stratum estimand strategy can be considered when the population of interest is the group of patients who would be in stratum $A=a$ if they received the experimental treatment, i.e., ${\cal E}_{Oa} \cup {\cal E}_{Ma}$ in the experimental treatment arm and ${\cal C}_{Oa} \cup {\cal C}_{Ma}$ in the control arm. 

The population-level summary is defined as 
\begin{equation}
\theta_a=G\left\{F_{a1}(\cdot), F_{a0}(\cdot)\right\}, a=1, \cdots, J,\label{eq:estimate}
\end{equation}   
in which $G(\cdot, \cdot)$ is an appropriate functional of two cumulative distribution functions, and $F_{aj}(t)$, $j=0,1$ is the cumulative distribution function of the conditional distribution $T^{(j)}|A=a$. $\theta_a$ could, for example, be the rate/mean difference for binary/continuous endpoints, or the hazard ratio/restricted mean survival time difference for time-to-event endpoints.

\subsection{Estimation}\label{ss:Estimation}

To target the specified estimand of interest in Section \ref{ss:estimand}, we propose the following novel weight-based approaches. 
Stratum status $A$ is fully observed in patients in the treatment arm with $M=0$, and a unit weight is assigned to patient from ${\cal E}_{Oa}$. Weight 
\begin{equation}
w_{a0}(x)=P(A=a|X=x) \label{eq:w0a}
\end{equation}
is assigned to patients in the control group based on the covariates at baseline/landmark.

Intuitively, a weighting approach for patients in the experimental treatment arm with $M=1$ analogous to the weight for the control arm based on baseline variables $X$ is expected to yield valid estimates for the treatment effect. However, such an approach requires that the efficacy outcome in the experimental arm, $T^{(1)},$ is conditionally independent of stratum $A$, given baseline variables $X$, which is the question of interest in this paper. Additional information observable only in the treatment arm, $B$, can help to identify the landmark status $A=a$ for patients with $M=1$, i.e. patients in ${\cal E}_{Ma}$. Here, $B$ is fully observed.

Post-baseline ADA measurements in our motivated study were longitudinally collected during the treatment cycles. The measures after the selected landmark time point, $B$, can be used to identify patients in ${\cal E}_{Ma}$. In many studies, patients with missing ADA status at the landmark time point receive ADA measurements after the landmark. Without loss of generality, $B$ could be the next available measurement after the landmark. 

Let 
\begin{equation}
w_{a1}(x,b)=P(A=a|X=x,B=b,R=1,M=0) \label{eq:w1a}
\end{equation}
be the weight for the treatment arm patients with missing stratum $A$, i.e. patients with $R=1$ and $M=1$. To ensure model for $A$ given $(X,B)$ and model for $A$ given $X$ only are compatible, $w_{a0}(.)$ can be estimated based on the following,  
\begin{equation}
w_{a0}(x)=\int_{\cal{B}} P(A=a|X=x,B=b,R=1,M=0)dF_B(b \mid X=x,R=1), \label{eq:w0a2}
\end{equation}
in which $\cal {B}$ is the support of $B$ and $F_B(\cdot\mid X, R)$ is the cumulative distribution function of $B\mid X, R.$ Any classification models can be used to estimate the above conditional probability.
The detailed steps for the proposed method are illustrated using logistic regression models as follows:
\begin{itemize}
	\item Step 1: Fit a logistic regression model with outcome $A$ on $B$ and baseline covariate $X$ based on patients in ${\cal E}_O={\cal E}_{O1} \cup \cdots \cup {\cal E}_{OJ}$.
	\item Step 2: Use the model in step 1 to compute the probability of $A=a$ for patients in the experimental arm with missing stratum status $({\cal E}_M={\cal E}_{M1} \cup \cdots \cup {\cal E}_{MJ})$, i.e., $P(A=a \mid X=x, B=b, R=1, M=1)$, denoting the estimated probability by $w_{a1}(x,b)$.
	\item Step 3: Fit a logistic regression model with outcome $B$ on $X$ based on all patients in the experiment arm $({\cal E}_{M1} \cup \cdots \cup {\cal E}_{MJ} \cup {\cal E}_{O1} \cup \cdots \cup {\cal E}_{OJ}).$
	\item Step 4: Use the models in step 1 and 3, plug in the baseline covariate $X$ for every patient in the control arm, and estimate $w_{a0}(x)=P(A=a \mid X=x, R=0)$ from $\int_{\cal{B}} \hat{P}(A=a|X=x,B=b)d\hat{F}_B(b \mid X=x)$, where $\hat{P}$ and $\hat{F}$ are estimated probability of $A$ given $(X,B)$ and cumulative distribution function of $B$ given $X$ from step 1 and 3 respectively. For a binary variable $B$ with values of $0$ and $1$, this simply means $w_{a0}(x)=\sum_{b=0,1}P(A=a|X=x,B=b)P(B=b|X=x)$
	\item Step 5: Assign weight 1 to patients from ${\cal E}_{Oa}$, weight $w_{a1}(x,b)$ to patients from ${\cal E}_{M},$ and weight $w_{a0}(x)$ to patients from the control arm.
\end{itemize}
Afterwards, the treatment effect $\theta_a$ can be estimated based on the distribution of outcomes among  patients from ${\cal E}_{Oa}$ and patients in the experimental arm with missing landmark status, i.e., ${\cal E}_{M},$ weighted by the weight function $w_{a1}(x,b)$, denoted by $\tilde{F}_{a1}(\cdot)$; and the distribution of outcomes among all control patients weighted by the weight function $w_{a0}(x)$, denoted by $\tilde{F}_{a0}(\cdot).$ The statistical inference for the estimated treatment effect can be made using nonparametric bootstrap method, where Steps 1-5 and the corresponding estimation of the treatment effect are repeated with generated bootstrapped samples.  In Appendix A, we demonstrate that $\tilde{F}_{a0}(t) = F_{a0}(t)$ and  $\tilde{F}_{a1}(t) = F_{a1}(t)$. To this end, the following assumptions are needed:

Assumption 1. The treatment assignment is independent of any post-baseline measures and baseline variables
\begin{equation}
R \perp (A,B,X) \label{assumption1} 
\end{equation}

Assumption 2. Principal ignorability in the control arm
\begin{equation}
T^{(0)} \perp A \,| \, X, \label{assumption2}
\end{equation}

Assumption 3. Ignorable missing and principal ignorability in the experimental treatment arm
\begin{equation}
M \perp A \,| \, X, B, \label{assumption3a}
\end{equation}
and
\begin{equation}
T^{(1)} \perp A \,| \, X, B, M=1. \label{assumption3b}
\end{equation}
Assumption 1 holds true in RCTs, when the difference between landmark and intend-to-treat populations is small. That is, the proportion of subjects who had event or censored before the landmark time point is small. This assumption is reasonable when an early landmark time point is chosen, so that the number of patients experiencing the clinical event of interest or early censoring prior to the landmark time point is small. Assumption 2 is required for the identifiability of $F_{a0}(\cdot)$ based on $(T^{(0)}, X)$ from the control arm. $A$ and $T^{(0)}$ can never be jointly observed as they are potential outcomes in ``parallel universes". A comprehensive list of variables that may explain the effect of $A$ on the clinical outcome $T^{(0)}$ should be included in the analysis. Assumptions 1 and 2 were also required in \citet{Bornkamp:2019}, where stratum $A$ was fully observed. Assumption 3, with its two parts (\ref{assumption3a}) and (\ref{assumption3b}), is required for addressing missing data in the treatment arm. Specifically, (\ref{assumption3a}) ensures the identifiability of the conditional distribution of $A \mid X, B$, which, coupled with (\ref{assumption3b}), ensures the identifiability of $F_{a1}(\cdot)$. Assumption $M \perp A \,| \, X, B$ is also needed for the estimation of the conditional distribution of $A\mid X$. A simple sufficient condition for Assumption 3 is 
\begin{equation}
T^{(1)} \perp A \perp M \mid  X, B.
\end{equation}
As an example, Assumptions 2 and 3 are satisfied in the following causal diagram

\begin{center}
\begin{tikzpicture}
\matrix (m) [matrix of math nodes,row sep=3em,column sep=4em,minimum width=5em]
{
T^{(0)}	& M &A \\
	 &X & \\
T^{(1)}	&  &B\\};
\path[-stealth]
(m-2-2) edge node [left] {} (m-1-1)
(m-2-2) edge node [left] {} (m-1-2)
(m-2-2) edge node [left] {} (m-1-3)
(m-2-2) edge node [left] {} (m-3-1)
(m-2-2) edge node [left] {} (m-3-3)
(m-1-3) edge node [left] {} (m-3-3)
(m-3-3) edge node [left] {} (m-1-3);
\end{tikzpicture}
\end{center}
\paragraph{Remark 1}{
In the discussion above, $B$ is considered as the next available measurement after the landmark time point. It may also represent other measurements only available in the treatment arm. Its inclusion makes Assumption 3 more likely to hold. However, in order to include general vector $B$, the model used in step 3 needs to be adjusted to allow consistent estimation of the conditional distribution of $B\mid X$ in the experimental arm.
}
\paragraph{Remark 2}{
It is possible to use other assumptions to replace Assumption 3, which has an overall goal of allowing consistent estimation of both $T^{(1)}\mid A$ and $A\mid X$ based on $(X, B)$ in the presence of missing data.  For example,  one may assume missing at random $M \perp A \mid T^{(1)}, X, B,$ and a parametric model for the conditional distribution of $(T^{(1)}, A)\mid X, B.$ 
}

The above framework is general, and the results are applicable to endpoints including mean/rate difference and time-to-event outcome. In the survival setting, 
$T=(Y, \Delta),$ $Y$ is the minimum of the survival and censoring times, and $\Delta$ is the censoring indicator. Let $Y^{(1)}$ and $Y^{(0)}$, as well as $\Delta^{(1)}$ and $\Delta^{(0)}$, be the corresponding potential outcomes in the experimental treatment and control arms, respectively.  We may let the previously defined functional $G(\cdot, \cdot)$ be the hazard ratio from fitting the Cox regression model as the treatment effect measure because this hazard ratio can be viewed as a functional of the distributions of $(Y^{(1)}, \Delta^{(1)})$ and $(Y^{(0)}, \Delta^{(0)})$. Note that this statement is true even if the proportional hazards assumption is violated, since the hazard ratio estimate from Cox regression still converges to a limit value dictated by the joint distribution of $(Y^{(g)}, \Delta^{(g)}, g=0, 1)$ \citep{lin1989robust}.  The downside of such a case is that the resulting hazard ratio estimator may not have a clear clinically meaningful interpretation.

If stratum status $A$ is fully observed for all experimental treatment arm patients (i.e., $P(M=0 \mid R=1)=1$), and the population-level summary is the difference in survival probabilities at a
given time point, with binary post-baseline subgroups (i.e. $J=2$), the proposed weight approach degenerates to the ``Weight Placebo Patients'' approach for time-to-event endpoint presented in \cite{Bornkamp:2019}. Similar assumptions are needed for our study, including (a) independent treatment assignment $R \perp (A, X)$ and (b) principal ignorability, i.e. all variables  that affect both $A$ and $T^{(0)}$ are measured and contained in the covariate vector $X:$ $T^{(0)} \perp A \mid X$.

Frequently it may not be possible to assess whether the weight model has been correctly specified.  Thus, rather than assessing the accuracy of the specification of the weight model, it is important to determine if weighting by the estimated propensity score induces a balance in measured covariates between treated and control patients \citep{Austin:2015}. This balance checking model diagnostic will be illustrated in Section \ref{sec_ada}.  

\section{SIMULATION}\label{sec_simulation}

Simulations were conducted to investigate the finite sample performance of the proposed method. Both binary and time-to-event outcomes with a binary stratum $A$ (positive or negative) are considered.  Sample size is set to 300, 600, 1000 or 2000 with 1:1 randomization. 

Simulation data are generated where the assumptions in Section \ref{ss:Estimation} are satisfied. For each individual, two independent baseline covariates $X_1$ and $X_2$ are generated from a standard multivariate normal distribution. $X_1$ and $X_2$ are assumed to be prognostically related to the clinical outcome and associated with the stratum status $A$ at the landmark time point. The next available stratum status after the landmark time point, $B$, is assumed to be positively correlated with $A$. This is reasonable because an individual positive at the landmark time point also has a high likelihood to be positive at the following  assessment and vice versa. $A$ is sampled from a Bernoulli distribution with
\begin{equation}
logit P(A=1|X, B)=\alpha_0+\alpha_1 X_1+\alpha_2 X_2+\alpha_3 B, 
\end{equation}    
where $(\alpha_0,\alpha_1,\alpha_2,\alpha_3)=(-2,1,-2,2)$. $B$ is sampled from a Bernoulli distribution with
\begin{equation}
logit P(B=1|X)=\gamma_0+\gamma_1 X_1 +\gamma_2 X_2,
\end{equation} 
where $(\gamma_0,\gamma_1,\gamma_2)=(-1,1,1)$.  Note that these two models imply that 

\begin{eqnarray*}
P(A=1 \mid X)& = & \frac{e^{\alpha_0+\alpha_1X_1+\alpha_2X_2}}{1+e^{\alpha_0+\alpha_1X_1+\alpha_2X_2}}\times \frac{1}{1+e^{\gamma_0+\gamma_1X_1+\gamma_2X_2}}+\\
& &\frac{e^{\alpha_0+\alpha_3+\alpha_1X_1+\alpha_2X_2}}{1+e^{\alpha_0+\alpha_3+\alpha_1X_1+\alpha_2X_2}}\times \frac{e^{\gamma_0+\gamma_1X_1+\gamma_2X_2}}{1+e^{\gamma_0+\gamma_1X_1+\gamma_2X_2}}
\end{eqnarray*}

The missing mechanism $M$ for landmark status $A$ is assumed to depend only on the baseline covariates $X_1$ and $X_2$  in the experimental treatment arm and is generated from a Bernoulli distribution with  
\begin{equation}
logit P(M=1|X)=\xi_0+\xi_1 X_1+\xi_2 X_2,
\end{equation} 
for which $(\xi_0,\xi_1,\xi_2)=(-2,-1,-3).$ The proportion of individuals with a missing value for $A$ is $30\%$ to allow a meaningful performance assessment of the approach in a realistic setting.  

For the binary outcome, data are generated from the following Bernoulli distribution with
\begin{align}
    logit P(Y^{(0)}=&1|X)=\beta_{00}+\beta_{1} X_1+\beta_{2} X_2,\\
    logit P(Y^{(1)}=&1|X,B)=\beta_{01}+\beta_{1} X_1+\beta_{2} X_2+\beta_{3} B,
\end{align}
for which $(\beta_{00},\beta_{01},\beta_{1},\beta_{2},\beta_{3})=(-2,2,1,2,-4)$.  For binary responses, the rate difference is used to summarize the treatment effect in each stratum. 

For time to event outcome, the outcomes are generated from 
\begin{align}
T^{(0)}\mid X \sim & \text{Exponential} \left\{\exp(\beta_{00}+\beta_{1} X_1 +\beta_{2} X_2)\right\}\\
T^{(1)}\mid X, B \sim & \text{Exponential}\left\{\exp(\beta_{01}+\beta_{1} X_1 +\beta_{2} X_2+\beta_{3} B)\right\}
\end{align}    
for which $(\beta_{00},\beta_{01},\beta_{1},\beta_{2},\beta_{3})=(-2,-3.5,1,3,4)$.
A fixed censoring time is selected to yield an administrative censoring rate of $20\%$. Hazard ratio is used to summarize the treatment effect in each stratum. Since the proportional hazards assumption is violated within stratum, the true hazard ratio is calculated from fitting the Cox regression based on all simulated patients, where $A$ in both treatment and control arms is known. The proposed weight-based analysis is conducted based on $(X_1, X_2, B)$. 

A set of sensitivity analyses are conducted. First, to investigate the performance of the approach with respect to inclusion of baseline covariates that are not related to clinical outcome or stratum status, noise baseline covariates $(Z_1, Z_2, Z_3)$ are generated from a standard multivariate normal distribution independently of all of the above variables. Coupled with $(X_1, X_2)$ and $B$, $(Z_1, Z_2, Z_3)$ are also used in estimating the weight, i.e, $X=(X_1,X_2, Z_1, Z_2, Z_3)'$ (scenario: weight including noise covariates). Furthermore, deviation from the assumption that all baseline covariates are collected that affect both efficacy $(T^{(0)},T^{(1)})$ and $(A,B)$ is assessed by including only $X_1$ and $B$ for computing the weight in the proposed approach, i.e., $X=X_1,$ (scenario: no principal ignorability). Finally, the impact of discarding the next available biomarker status $B$ in the weight computation for patients in the treatment arm with missing $A$ is investigated using the usual weight for control arm patients (\ref{eq:w0a}) but using a weight based only on $(X_1,X_2)$ for patients with $R=1$ and $A$ missing (scenario: weight without $B$). The weighting scheme without $B$ is problematic in estimating both the conditional probability $P(A=1|X_1, X_2)$ and $F_{a1}(\cdot).$ The failure in estimating the former is due to the misspecification of the logistic regression, and the latter is due to the dependence of $T^{(1)}$ on $B$ and thus $A,$ even after adjustment for $X_1$ and $X_2.$ For each of the above scenarios, $500$ simulations were completed, and the results are reported in Table \ref{Tab_simu_binary} and Table \ref{Tab_simu_os}. Standard error (SE) from the simulation, average SE estimates (SEEs) from 1000 samples generated by non-parametric bootstrapping and the estimated coverage probability of the 95\% confidence intervals (95\% CP) are tabulated. 

\begin{table}[H]
	\caption{Simulation summaries for binary outcome}
	\label{Tab_simu_binary}
	\begin{center}
		\footnotesize
		\renewcommand\arraystretch{1.5}
		\renewcommand\tabcolsep{2pt}
		\begin{tabular}{ccccccccccccc} \\
			\hline
			\hline
			&  & \multicolumn{2}{c} {$n=300$} && \multicolumn{2}{c} {$n=600$} &&\multicolumn{2}{c} {$n=1000$} &&\multicolumn{2}{c} {$n=2000$}\\
			\cline{3-4}\cline{6-7}\cline{9-10}\cline{12-13}
			&& $A=1$ & $A=0$ && $A=1$ & $A=0$&&$A=1$ & $A=0$ &&$A=1$ & $A=0$\\
			\hline
			True &Response difference & 25.6\% & 40.9\% && 25.8\% & 41.2\% &&25.7\% & 41.2\% && 25.8\% & 41.2\%\\
			\hline
			Proposed method & Mean & 25.4\% & 40.9\%& & 26.1\% & 41.0\% &&25.8\% & 41.2\% && 25.7\% & 41.2\%\\
			& SE & 0.081 & 0.062 & & 0.056 & 0.042 &&0.043 & 0.033 && 0.030 &0.023\\
			&$\mbox{SEE}^{a}$& 0.080 & 0.060& & 0.056 & 0.042 &&0.043 & 0.033 && 0.030 &0.023\\
			& 95\% $\mbox{CP}^{b}$& 94.0\% & 94.8\%& & 95.0\% &95.0\% &&95.4\% &94.6\% && 94.2\% & 95.8\%\\
			\hline
			Weight including noise covariates & Mean &25.4\% &40.9\%& & 26.1 \% & 41.0\% &&25.8 \% & 41.2\% && 25.7\% & 41.3\%\\
			&SE & 0.083 & 0.062& & 0.056 & 0.042 &&0.043 & 0.033 &&0.030&0.023\\
			& $\mbox{SEE}^{a}$ & 0.083 & 0.061& & 0.056 & 0.042 &&0.043 & 0.033 && 0.030 &0.023\\
			& 95\% $\mbox{CP}^{b}$ & 94.0\% & 95.4\%& & 95.8\% & 94.6\% &&95.6\% & 94.6\% &&94.0\% & 96.2\%\\
			\hline
			No principal ignorability & Mean &15.3\% & 42.0\% & & 16.3\% & 42.1\% &&16.3\% & 42.1\% &&15.9\% & 42.2\%\\
			&SE &0.101 & 0.056& & 0.069 & 0.037 &&0.053 & 0.030 && 0.038 &0.021\\
			& $\mbox{SEE}^{a}$ &0.097 & 0.053& & 0.068 & 0.037 &&0.039 & 0.053 && 0.037 &0.020\\
			& 95\% $\mbox{CP}^{b}$ & 80.4\% & 93.4\% && 71.0\% & 93.8\% &&56.6\% & 91.8\% && 26.0\% &90.4\%\\
			\hline
			Weight without $B$ & Mean & 27.4\%& 39.9\% & &28.0\% &40.0\% &&27.8\% &40.1\% &&27.6\% &40.2\%\\
			&SE & 0.080 & 0.061 & &0.055 & 0.041&&0.042 & 0.032 &&0.030 &0.023\\
			&$\mbox{SEE}^{a}$ & 0.079 & 0.059& &0.055 & 0.041 &&0.042 & 0.032&&0.030 &0.023\\
			&95\% $\mbox{CP}^{b}$ & 94.8\% &94.6\% && 92.6\% & 94.6\% &&92.6\% & 93.6\% && 91.0\% &92.6\%\\
			\hline
			\hline
		\end{tabular}
		\footnotesize{Explanations:$^{a}$Average SE estimates from 1000 bootstrap; $^{b}$ estimated coverage probability of the 95\% confidence interval}\\
	\end{center}
\end{table}

\begin{table}[H]
	\caption{Simulation summaries for time-to-event outcome}
	\label{Tab_simu_os}
	\begin{center}
		\footnotesize
		\renewcommand\arraystretch{1.5}
		\renewcommand\tabcolsep{2pt}
		\begin{tabular}{ccccccccccccc} \\
			\hline
			\hline
			&  & \multicolumn{2}{c} {$n=300$} && \multicolumn{2}{c} {$n=600$} &&\multicolumn{2}{c} {$n=1000$} &&\multicolumn{2}{c} {$n=2000$}\\
			\cline{3-4}\cline{6-7}\cline{9-10}\cline{12-13}
			&& $A=1$ & $A=0$ && $A=1$ & $A=0$&&$A=1$ & $A=0$ &&$A=1$ & $A=0$\\
			\hline
			True & log (HR) & -0.107 & -0.311&& -0.087 & -0.306 &&-0.094 & -0.311 && -0.095 &-0.310\\
			\hline
			Proposed method & Mean& -0.102 &-0.315&  & -0.093 & -0.304 && -0.092 & -0.310 && -0.096 &0.309\\
			& SE & 0.217 & 0.133 & & 0.157 & 0.098 &&0.120 & 0.070 && 0.082 & 0.053\\
			&$\mbox{SEE}^{a}$ & 0.204 & 0.127 && 0.140 & 0.089 &&0.109 & 0.068 && 0.076 & 0.048\\
			& 95\% $\mbox{CP}^{b}$ & 92.2\% & 95.0\% && 93.4\% &92.0\% &&92.4\% &94.2\% && 92.6\% & 94.0\%\\
			\hline
			Weight including noise covariates & Mean &-0.103 &-0.316 & &  -0.094 & -0.304 && -0.092 & -0.309 && -0.096 &-0.309\\
			&SE & 0.219 & 0.134& & 0.156 & 0.098 &&0.121 & 0.070 && 0.082 &0.053\\
			& $\mbox{SEE}^{a}$ & 0.211 &0.131& & 0.142 & 0.090 &&0.110 & 0.069 && 0.076 & 0.048\\
			& 95\% $\mbox{CP}^{b}$ & 93.8\% & 95.0\% && 93.2\% & 92.0\% &&92.2\% & 94.6\% && 92.6\% &93.8\%\\
			\hline
			No principal ignorability & Mean &-0.216 & -0.233& & -0.201 & -0.223 && -0.200 & -0.227 && -0.204 &-0.226\\
			&SE & 0.230 & 0.140 && 0.162 & 0.104 &&0.128 & 0.076 && 0.091 &0.054\\
			& $\mbox{SEE}^{a}$ &0.231 &0.125 & & 0.160 & 0.088 &&0.124 & 0.068 && 0.087 &0.048\\
			& 95\% $\mbox{CP}^{b}$ &91.6\% &87.0\% & & 81.8\% & 70.4\% &&83.6\% & 75.6\% && 76.8\% & 59.8\%\\
			\hline
			Weight without $B$ & Mean &-0.146 & -0.288 & &-0.137 &-0.278 && -0.133 & -0.284 && -0.138 &-0.283\\
			&SE & 0.212 & 0.131&& 0.151 & 0.098&&0.117 & 0.070 && 0.080 & 0.053\\
			&$\mbox{SEE}^{a}$ & 0.194 & 0.126 && 0.134 & 0.088 &&0.104 & 0.068 && 0.073 &0.048\\
			&95\% $\mbox{CP}^{b}$ & 91.6\% &94.2\%&& 91.0\% & 90.6\% &&90.2\% & 93.0\% && 89.2\% &87.8\%\\
			\hline
			\hline
		\end{tabular}
			\footnotesize{Explanations:$^{a}$Average SE estimates from 1000 bootstrap; $^{b}$ estimated coverage probability of the 95\% confidence interval}\\
	\end{center}
\end{table}

The results suggest that the biases of the proposed estimators are minimal, and the bootstrap method performs well in estimating the variance, yielding estimates of variance similar to their empirical counterparts, and confidence intervals with a reasonable coverage level. The weighting method based on $X=(X_1, X_2, Z_1, Z_2, Z_3)'$ and $B$ also yields unbiased estimates and confidence intervals with a reasonable coverage probability.  The weighting approach missing $X_2$ or $B$ resulted in substantial biases in estimating the stratum-specific treatment effect. Without $B,$ the estimated treatment effects for $A=1$ and $A=0$ become more similar, probably due to the reduced ability to differentiate between patients with $A=1$ from those with $A=0$ in the treatment arm without using the information in $B.$ 

\section{APPLICATION}\label{sec_ada}

\subsection{Assessment of Anti-Drug Antibodies (ADA)} \label{s:ice}

For a biologic therapeutic, characterization of immunogenicity is an important part of the drug development process. This includes the investigation of ADA formation and their potential impact on the long-term outcome of the biologic therapeutic \citep{DeGroot:2007,Baker:2010}. In clinical trials with a biologic therapeutic that is administered multiple times to a patient, ADAs are measured at a carefully planned series of time points, including an ADA assessment at baseline. With industry standard definitions for determining ADA status \citep{Shankar:2014}, patients are considered ADA-negative if they are negative at all ADA assessments after treatment initiation. Patients are also ADA-negative if they are ADA-positive at baseline but the ADA titer (concentration) does not increase by more than a pre-specified value relative to baseline; a status also referred to as treatment-unaffected ADA-negative. Patients are considered ADA-positive if they test ADA-positive at least once after treatment initiation. The only exception are patients who are treatment-unaffected ADA-negative; these patients are considered ADA-negative. 
    
Hence, by definition, ADA status in an RCT is a post-randomization variable induced by treatment with the potential to impact the interpretation of the clinical outcome. It can thus be considered an intercurrent event in the language of the ICH E9 addendum \citep{ICHE9:2019}. 

The approach introduced in this paper was applied to ADAs in the pivotal IMpower150 trial. IMpower150 was a randomized and controlled Phase 3 study investigating the CIT atezolizumab in chemotherapy-naïve patients with metastatic non-squamous non-small cell lung cancer \citep{Socinski:2018}. CITs are biologic treatments that boost the body's natural defenses to fight cancer \citep{Dillman:2011}. The application focused on arm B and arm C as described below which showed an overall survival and progression-free survival benefit of arm B over arm C, resulting in various regulatory approvals world wide.    

The treatment regimens of interest were as follows:
\begin{itemize}
	\item Arm B: ABCP ($n=400$): atezolizumab+bevacizumab+carboplatin+paclitaxel induction (four or six 21-day cycles) followed by
	atezolizumab+bevacizumab maintenance (21-day cycles)
	\item Arm C: BCP ($n=400$): bevacizumab+carboplatin+paclitaxel induction (four or six 21-day cycles) followed by bevacizumab
	maintenance (21-day cycles)
\end{itemize}

To investigate a potential impact of ADA on outcome, the method described earlier was applied to the principal stratum formed by ADA as illustrated in Table \ref{tableA} with $J=2$. The specific scientific questions on the principal stratum were: (i) was the treatment effect clinically meaningful for ADA-positive and ADA-negative patients and (ii) did the treatment effect differ between ADA-positive and ADA-negative patients, i.e. does ADA formation affect the treatment benefit of atezolizumab. Because atezolizumab was administered in the experimental treatment arm but not in the control arm, the potential for ADA formation is unknown in the control arm. The percentage of ADA-positive patients in treatment arm, i.e., arm B, was $36\%$. 

ADA is an internal time-varying variable that is only meaningful for survivors (see Section \ref{ss:estimand}). In particular, there is the potential for an immortal (event-immune) bias due to the requirement for patients to live long enough to allow a post-baseline ADA assessment \citep{Lash:2009}. Furthermore, patients with longer follow-up have a higher chance to be ADA-positive due to multiple ADA assessments. To address these limitations, a landmark approach was applied so that patients were considered to have (landmark) ADA-positive status if, and only if, they were ADA-positive before the landmark time point. A very small group of patients who had an event or were censored before the landmark time point were excluded from this analysis. Based on general principles given for landmark selection in Appendix B, the landmark time point in this application was set at $4$ weeks after the baseline visit. The first post-baseline scheduled ADA sampling time point was at week 3, and an additional week was added to capture ADA assessments from a nominal week 3 study visit that was slight delayed relative to the planned testing schedule. Approximately 11.3\% of the ADA stratum information at the landmark time point in arm B was missing. 

A set of baseline covariates that are prognostically related to long-term outcome in lung cancer and considered reasonably likely to be associated with ADA status were selected. Specifically, a set of $15$ baseline covariates were ascertained based on clinical and statistical considerations, including aggregation of expert opinions, an assessment of previous clinical data, and a careful literature review. The baseline covariates were pre-specified before the statistical analysis. Baseline covariates that are prognostically associated with the long-term outcome but not associated with the ADA status would not introduce bias.

\subsection{RESULTS}

\subsubsection{Main results}

An examination of the distribution of the $15$ baseline covariates revealed covariate distribution differences between ADA-positive and ADA-negative patients in the experimental arm, prognostically favoring the ADA-negative patients at the week $4$ landmark time point. The proportion of patients either had events or censored before the landmark time point was $3.4\%$.

\begin{figure}
	\centering
	\includegraphics[width=0.7\linewidth]{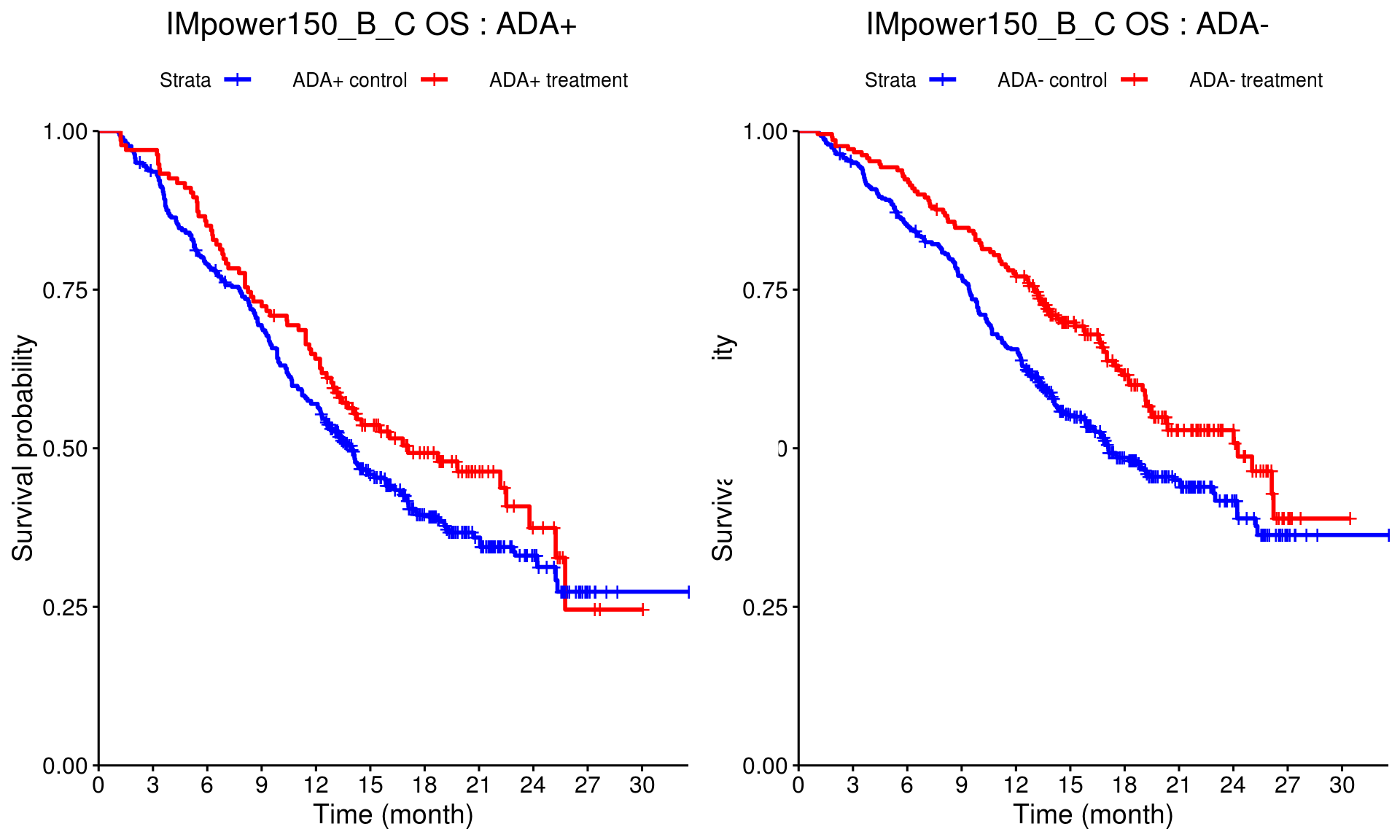}
  \caption{Kaplan-Meier plot by week 4 ADA landmark}
  \label{fig:kmadaweek4}
\end{figure}

The proposed weight approach for the week 4 landmark yielded a hazard ratio of 0.766 (95\%CI: 0.574, 1.021) for the ADA-positive stratum and 0.732 (95\%CI: 0.578, 0.926)  for the ADA-negative stratum. Hence, the hazard ratio point estimates were similar between the ADA strata and the 95\% CIs overlapped heavily. The corresponding weighted Kaplan-Meier curves for ADA-positive and ADA-negative strata are presented in Figure \ref{fig:kmadaweek4}, demonstrating sustained separation between the curves for each of the ADA groups and the corresponding control curves.  

\begin{figure}
	\centering
	\includegraphics[width=1\linewidth]{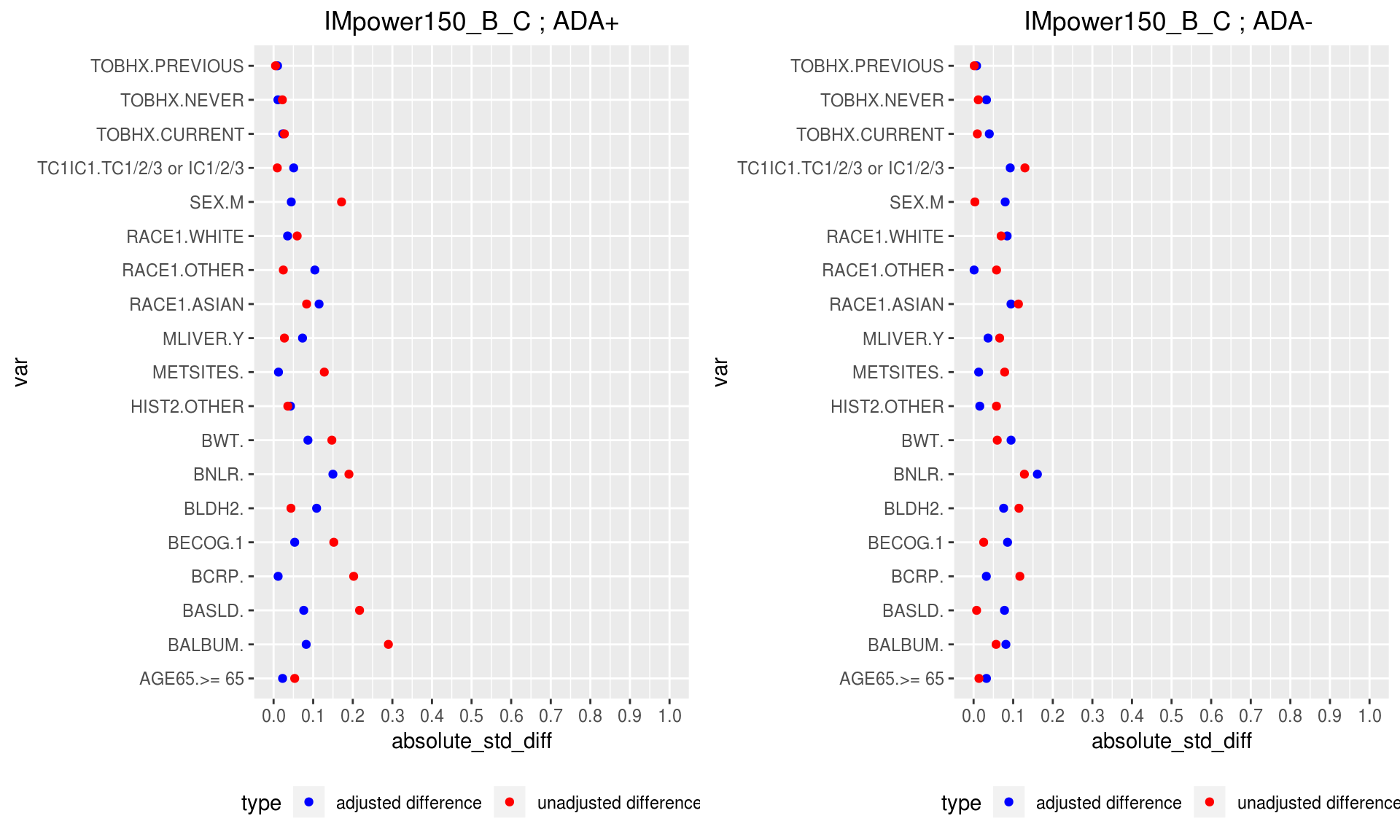}
  \caption{AMSD plot by week 4 ADA landmark. Blue dots show the ASMD per covariate when comparing each ADA stratum with its adjusted (weighted) control. Red dots show the ASMD per covariate when comparing each ADA stratum with the unadjusted (overall unweighted) control. Details on the covariates are provided in Appendix C.}
  \label{fig:amsd_week4}
\end{figure}

Results of standard model diagnostics based on the absolute standardized mean difference (ASMD) are presented in Figure \ref{fig:amsd_week4}. In addition, given the small sample size in the ADA strata, Table \ref{Tab_amsd} summarizes the number of expected (from a theoretical RCT with the same sample size, see Appendix B) and observed covariates with an ASMD that is greater than the defined threshold. The observed numbers of covariates with adjusted ASMD values above 0.1 and 0.25 were both in line with the expected number in an RCT and smaller than those before adjustment. Overall, covariates were well balanced between two arms after proposed weighting.

\begin{table}[H]
	\caption{ASMD benchmarking vs RCT by week 4 ADA landmark}
	\label{Tab_amsd}
	\begin{center}
		\footnotesize
		\renewcommand\arraystretch{1.5}
		\renewcommand\tabcolsep{2pt}
		\begin{tabular}{cccccccc} \\
			\hline
			\hline
			&\multicolumn{2}{c}{No. of  ASMD $> 0.1$}&&\multicolumn{2}{c}{No. of ASMD $> 0.25$}\\
			\hline
			Stratum & {\thead{Expected}}& {\thead{Observed}} && {\thead{Expected}} & {\thead{Observed}}\\
			\hline
			ADA+  & 6.9 & 3 && 1 & 0\\
			ADA-  & 4.2 & 2 && 0.1 & 0\\
			\hline
			\hline
			
		\end{tabular}
	\end{center}
\end{table}

\subsubsection{Sensitivity analyses: alternative strategies to handle missing ADA results}\label{section:ada:support}

Two extreme imputations of the missing landmark status were considered in a sensitivity analysis: negative imputation considered all patients with a missing ADA status at the landmark to be ADA-negative, and positive imputation considered all patients with a missing ADA status at the landmark to be ADA-positive.  
Using the nomenclature from Table \ref{tableA}, the negative imputation method targets the estimand ${\cal E}_{O1}\cup{\cal E}_{M1}\cup{\cal E}_{M2}$ vs. ${\cal C}_{O1}\cup{\cal C}_{M1}\cup{\cal C}_{M2}$ and ${\cal E}_{O2}$ vs. ${\cal C}_{O2}$, while the positive imputation method targets the estimand ${\cal E}_{O1}$ vs. ${\cal C}_{O1}$ and ${\cal E}_{O2}\cup{\cal E}_{M1}\cup{\cal E}_{M2}$ vs. ${\cal C}_{O2}\cup{\cal C}_{M1}\cup{\cal C}_{M2}$.  
With all of the missing landmark status imputed, the proposed approach degenerates to the weight approach proposed in \cite{Bornkamp:2019}. Despite the methods being extremes from an imputation perspective, it is important to note that the efficacy results from  two methods above are not necessarily extremes in terms of estimated treatment effects. 
 
Another sensitivity analysis referred to as the ``complete case analysis'' targets the estimand  ${\cal E}_{O1}$ vs. ${\cal C}_{O1}$, ${\cal E}_{O2}$ vs. ${\cal C}_{O2}$ and ${\cal E}_{M1}\cup{\cal E}_{M2}$ vs. ${\cal C}_{M1}\cup{\cal C}_{M2}$, where the probability of conterfactural missing ADA landmark status in the control arm is estimated based on observed covariates.

\begin{figure}
	\centering
	\includegraphics[width=1\linewidth]{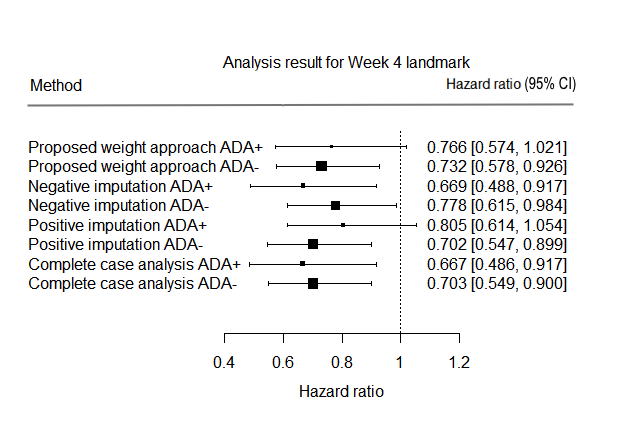}
  \caption{Alternative strategies to handle missing ADA}
  \label{fig:forest}
\end{figure}

The results of these three sets of analyses are summarized in Figure \ref{fig:forest}. The estimands for the ADA-negative stratum for positive imputation and complete case analysis are identical.  Similarly, the estimands for the ADA-positive stratum based upon negative imputation and  complete case analysis are identical. 
As shown in Figure \ref{fig:forest}, similar hazard ratios were observed when the same estimand was targeted. The estimated treatment effects in both ADA positive and ADA negative strata are fairly similar in all sensitivity analyses.

\section{DISCUSSION}

A weighted approach for estimating treatment effects in principal strata defined by an intercurrent event in a RCT was presented. The determination of such a treatment effect is challenging because of two elements. First, because one does not observe the outcomes of the patients in such a principal stratum would they have obtained the control treatment. Secondly, because an indicator (yes/no) for missing data in the intercurrent event is observed in the treatment arm but not in the control arm. Assumptions for drawing causal inference for such a setting were introduced. A simulation study and a real world application to estimating the treatment effect for principal strata defined by ADAs in an RCT showed that the approach performs well.  

Different estimands may be considered with missing $A$ in this problem. One example is imputing all the missing stratum $A$ with $A=a$ as illustrated in section \ref{section:ada:support} of the motivated example. The targeted estimand becomes ${\cal E}_{Oa}\cup{\cal E}_{M1}\cup\cdots\cup {\cal E}_{MJ}$ vs. ${\cal C}_{Oa}\cup{\cal C}_{M1}\cup\cdots\cup{\cal C}_{MJ}$. Another estimand could be considered is ${\cal E}_{Oa}$ vs. ${\cal C}_{Oa}$ if ``complete case analysis" is of interest. In the latter case, the defined target estimand indicates that the missing stratum is only handled on the level of the estimand and not on the level of estimation. The limitation is that the patients population corresponding to such an estimand may not be intepretable.   

Unobserved counterfactural principal stratum $A$ in the control group can be viewed as a missing data problem under the causal inference framework \citep{Rubin:1974}; hence, a popular method like multiple imputation could also be considered. The counterfactural stratum in the control arm and the missing stratum in the experimental treatment arm with $M=1$ can be imputed based on comprehensive modeling as the proposed approach.

\if0\blind
{\section*{Acknowledgments}
The authors would like to thank Nitzan Sternheim, Marcus Ballinger, Jane Ruppel and Benjamin Wu for important discussions on the biological and clinical aspects of immunogenicity. We would also like to thank Marcel Wolbers and  Kaspar Rufibach for valuable comments on the methodological part. Finally, we would like to thank Ning Leng, Heng Wang, and Joel Laxamana for supporting execution of the approach. 
} \fi

\section*{APPENDICES}
\subsection*{Appendix A: theoretical justification of the proposed weight approach}

First, we want to show that $\tilde{F}_{a0}(t) = F_{a0}(t)$. From Assumption 1,
\begin{align}
\tilde{F}_{a0}(t)&\propto \int_0^t\int_{{\cal X}} f(u|X=x, R=0)w_{a0}(x)dP(X\leq x|R=0) du \ \nonumber\\
&=\int_0^t\int_{{\cal X}} f(u|X=x, R=0)\left\{\int_{\cal B}P(A=a|X=x,B=b,M=0)dP(B\leq b|X=x)\right\} \nonumber\\
&~~\times \frac{dP(X\leq x)}{dP(X\leq x|A=a)}dP(X\leq x|R=0,A=a) du \ \nonumber \\
&=\int_0^t\int_{{\cal X}} f(u|X=x, R=0)\left\{\int_{\cal B}P(A=a|X=x,B=b)dP(B\leq b|X=x)\right\}\label{eqf101}\\
&~~\times \frac{P(A=a)}{P(A=a|X=x)} dP(X\leq x|R=0,A=a) du \ \nonumber \\
&\propto \int_0^t\int_{{\cal X}} f(u|X=x, R=0,A=a) P(A=a|X=x) dP(X\leq x|R=0,A=a) = F_{a0}(t),\label{eqf102}
\end{align}
for which $f(u|X=x, R=0)$ is the density function for the conditional distribution $T^{(0)}|X=x.$  Equation (\ref{eqf101}) stems from Assumption 3 (\ref{assumption3a}) and equation (\ref{eqf102}) holds from Assumption 2 (\ref{assumption2}). Therefore, $\tilde{F}_{a0}(t) = F_{a0}(t),$ because both are proper cumulative distribution functions.

Now we look at the treatment arm conditional on $B$ and missing indicator $M$,
\begin{align*}
F_{a1}(t)&=P(M=0| A=a, R=1)\int_0^t\int_{{\cal X}}\int_{{\cal B}}f(u|B=b, X=x,M=0, R=1,A=a)\\
&~~dP(B\leq b|X=x,M=0, R=1,A=a)\\
&~~dP(X\leq x|M=0, R=1,A=a)du\\
&+P(M=1|A=a, R=1)\int_0^t\int_{{\cal X}}\int_{{\cal B}}f(u|B=b, X=x, M=1, R=1,A=a)\\
&~~dP(B\leq b|X=x, M=1, R=1,A=a)\\
&~~dP(X\leq x|M=1, R=1,A=a)du.
\end{align*}

We also have
\begin{align*}
\tilde{F}_{a1}(t)& =\frac{P(M=0, A=a | R=1)}{P(A=a | R=1)}\int_0^t\int_{{\cal X}}\int_{{\cal B}}f(u|B=b, X=x,M=0, R=1,A=a)\\
&~~dP(B\leq b|X=x,M=0, R=1,A=a)\\
&~~dP(X\leq x|M=0, R=1,A=a)du\\
&+\frac{P(M=1|R=1)}{P(A=a | R=1)}\int_0^t\int_{{\cal X}} \int_{{\cal B}}f(u|B=b, X=x,M=1, R=1)\\
&~~P(A=a|B=b, X=x,M=0,R=1)dP(B\leq b|X=x, M=1, R=1)\\
&~~dP(X\leq x|M=1, R=1)du.\\
\end{align*}

$\tilde{F}_{a1}(t)$ is a proper cumulative distribution function. 
We only need to show that the second part of the above equation equals to the second part in $F_{a1}(t)$, i.e.
\begin{align*}
&P(M=1, A=a | R=1)\int_0^t\int_{{\cal X}}\int_{{\cal B}}f(u|B=b, X=x, M=1, R=1,A=a)\\
&~~dP(B\leq b|X=x, M=1, R=1,A=a)\\
&~~dP(X\leq x|M=1, R=1,A=a)du\\
&= P(M=1|R=1)\int_0^t\int_{{\cal X}} \int_{{\cal B}}f(u|B=b, X=x, M=1, R=1)\\
&~~P(A=a|B=b, X=x, M=0, R=1)dP(B\leq b|X=x, M=1, R=1)\\
&~~dP(X\leq x| M=1, R=1)du. 
\end{align*}

From equation (\ref{assumption3b}) in Assumption 3, 
\begin{align*}
LHS=&P(M=1, A=a | R=1)\int_0^t\int_{{\cal X}}\int_{{\cal B}}f(u|B=b, X=x, M=1, R=1)\\
&~~dP(B\leq b|X=x, M=1, R=1,A=a)\\
&~~dP(X\leq x|M=1, R=1,A=a)du\\
=& \int_0^t\int_{{\cal X}}\int_{{\cal B}}f(u|B=b, X=x, M=1, R=1) P(M=1, A=a | R=1)\\
&~~dP(B\leq b, X\leq x| M=1, R=1, A=a)du\\
=& \int_0^t\int_{{\cal X}}\int_{{\cal B}}f(u|B=b, X=x, M=1, R=1) dP(B\leq b, X\leq x, M=1,A=a | R=1)du.
\end{align*}

From equation (\ref{assumption3a}),
\begin{align*}
RHS=&P(M=1|R=1)\int_0^t\int_{{\cal X}} \int_{{\cal B}}f(u|B=b, X=x, M=1, R=1)\\
&~~P(A=a|B=b, X=x, M=1, R=1)dP(B\leq b|X=x, M=1, R=1)\\
&~~dP(X\leq x| M=1, R=1)du\\
=&P(M=1|R=1)\int_0^t\int_{{\cal X}} \int_{{\cal B}}f(u|B=b, X=x, M=1, R=1)\\
&~~dP(B\leq b, X\leq x,A=a | M=1, R=1)du\\
=& \int_0^t\int_{{\cal X}}\int_{{\cal B}}f(u|B=b, X=x, M=1, R=1) dP(B\leq b, X\leq x, M=1,A=a | R=1)du.
\end{align*}

Therefore, $\tilde{F}_{a1}(t)=F_{a1}(t).$

\subsection*{Appendix B: Practical Guidance}\label{subsec_LM_diagnostics}

\subsubsection*{Landmark Selection}

If $A$ carries mortality information, different biases can be introduced (Section \ref{s:ice}). To address those biases, a landmark analysis can be applied. In general, selection of a landmark time point to address an internal time-varying  will need to be done on a case-by-case basis. However, some practical guidance is needed to achieve a balance between maximizing the number of patients in a certain stratum $A=a$ at the landmark (e.g., ADA-positive stratum in the IMpower150 ADA example) and minimizing the number of patients excluded overall due to event or censoring prior to that landmark. Also, an early landmark time point is usually preferred so that the landmark analysis population is reasonably close to the overall population and Assumption 1 holds. An early landmark time point may also be preferable to enable clinical decision-making (e.g., on treatment patient selection). Finally, the assessment schedule for $A$ should be considered and the landmark selection should preferably allow for some small deviations from the planned assessments because patients may come into the care center for the assessment of $A$ a few days later for various reasons. This may be particularly relevant if the treatment is not administered very frequently (e.g., every 4 weeks) as in many cases assessment of $A$ will happen shortly prior to administration.

\subsubsection*{Model diagnostics}

Model diagnostics of the approach introduced in this paper include assessment of balancing of the selected baseline covariates. This can be done using the ASMD \citep{Austin:2015} as this method is not influenced by sample size and can be used to compare balancing in measured variables between each ADA subgroup in the experimental arm and its appropriate control when different weights are assigned to the same patient in the control group. In many practical applications of principal stratification in an RCT, one or multiple stratum sample sizes will be small. The literature suggests that for small sample sizes, despite efforts to modify the weight model, it may not be possible to have all estimated ASMD values below an arbitrary threshold such as 0.1 or 0.25 \citep{Austin:2009}.  Hence, for practical applications, the number of baseline covariates after adjustment observed to exceed certain thresholds (such as 0.1 and 0.25) can be compared with the expected number of covariates exceeding those thresholds in an RCT with sample sizes in both arms identical to the size of the corresponding ADA stratum (assuming 1:1 randomization).  The rationale for this approach is that results would be considered acceptable if they provided a balance close to what would be achieved in an RCT with the same (small) sample size.  In an RCT, the ASMD has an SE of $\sqrt{1/n_1+1/n_0}$, for which $n_1$ and $n_0$ are the sample sizes for experimental and control arms respectively.  This can then be used to compute the number of expected covariates above a certain threshold such as $0.1$ and $0.25$ in our application in Section \ref{sec_ada}.  

\subsection*{Appendix C: Variables included in the IMpower150 study}\label{variable_list}
\begin{table}[H]
   	\captionsetup{labelformat=empty}
	\caption{Table C.1: Covariates included in the IMpower150 study}\label{table_variable}
	\begin{center}
	\footnotesize
		\begin{tabular}{ |c|c| } 		
			\hline
			 Covariate (variable name)& Categorical (levels) versus Continuous Covariates\\
			\hline
			Age (AGE65) & Categorical ( $<65$ years vs.  $\geq 65$ years)\\
			Albumin (BALBUM) & Continuous\\ 
			Sum of longest diameters (BASLD) & Continuous\\
			C-reactive protein (BCRP) & Continuous\\
			ECOG performance status & Categorical (0 vs. 1)\\
			Lactate dehydrogenase (BLDH2) & Continuous\\
			Neutrophil-lymphocyte ratio (BNLR) & Continuous\\
			Weight (BWT) & Continuous\\
			Histology (HIST2) & Categorical \\
			& (adenocarcinoma or bronchioloalveolar carcinoma vs. other)\\
			Number of metastatic sites (METSITES) & Continuous\\
			Liver metastases (MLIVER) & Cateogrical (yes vs. no)\\
			Race (RACE) & Categorical (White vs. Asian vs. Other)\\
			Sex (SEX) & Categorical (male vs female)\\
			PD-L1 (TC1IC1) & Categorical (TC1/2/3 or IC1/2/3 vs. TC0 and IC0)\\
			Tobacco history (TOBHX) & Categorical (previous vs. current vs. never)\\
			\hline
		\end{tabular}
	\end{center}
\end{table}

\bibliographystyle{statinmed}

\bibliography{refs_MI}
\end{document}